\documentclass[runningheads,a4paper]{llncs}

\usepackage{amssymb}
\usepackage{graphicx}
\usepackage{caption}
\usepackage{subcaption}
\usepackage{amsmath}
\usepackage{subfig} 

\usepackage{url}

\newcommand{\DD}{\, \displaystyle}

\newcommand{\thmax}{\,\theta_{\textnormal{max}}}

\newcommand{\ttt}{\,\theta}
\newcommand{\lam}{\,2L(0)+N(0)}

\newcommand{\fracc}[2]{\, \displaystyle \frac{ #1}{ #2}}

\newcommand{\ave}[1]{\, \langle{#1}\rangle}

\newcommand{\morabba}[1]{\,\begin{flushright}
 \Rectsteel \\
\end{flushright}}

\newcommand{\CC}[2]{\, \binom{#1}{#2} 
}

\usepackage{setspace}

\newcommand{\al}[1]{\,\begin{align}
                   #1 
                   \end{align}
}
\newcommand{\all}[2]{\,\begin{align}
                   #1 
                    \label{#2}
                   \end{align}
}

\begin{document}

\mainmatter  

\title{Generalized Friendship Paradox: \\ 
 An Analytical Approach}

\titlerunning{Generalized Friendship Paradox: An Analytical Approach}

%
%
\author{Babak Fotouhi$^{1,2}$, Naghmeh Momeni$^{1}$ and  Michael G. Rabbat$^1$}
\authorrunning{Babak Fotouhi, Naghmeh Momeni and  Michael G. Rabbat}

\institute{$^1$ Department of Electrical and Computer Engineering\\
McGill University, Montr\'eal,  Canada\\
$^2$ Department of Sociology, 
McGill University, Montr\'eal,   Canada\\
\email{babak.fotouhi@mail.mcgill.ca}, 
\email{naghmeh.momenitaramsari@mail.mcgill.ca}, 
\email{michael.rabbat@mcgill.ca}}


%
%

\maketitle

\begin{abstract}
The friendship paradox refers to the sociological observation that,  while   the  people's   assessment of their own popularity  is typically self-aggrandizing, in reality they are less popular than their friends.  The generalized friendship paradox is the average alter superiority observed empirically in social settings,   scientific collaboration networks, as well as online social media. 
We posit a quality-based network growth model in which the chance for a node to receive new links depends both on its degree and a quality parameter. Nodes are assigned qualities the first time they join the network, and these do not change over time. We analyse the model theoretically, finding expressions for the joint degree-quality distribution and nearest-neighbor distribution. We then demonstrate that this model exhibits both the friendship paradox and the generalized friendship paradox at the network level, regardless of the distribution of qualities. We also show that, in the proposed model, the degree and quality of each node are positively correlated regardless of how node qualities are distributed.
\end{abstract}

\section{Introduction} \label{sec:intro}
 
The friendship paradox is a phenomenon observed in various social networks. The term was coined by Feld~\cite{Feld}. It has been empirically observed that people's perception of their own popularity is self-aggrandizing; most people believe that they are more popular than their friends on average~\cite{Zuck}. However,  Feld observed that in reality, most people have fewer friends than their friends do.  In~\cite{GFP_Fowler}, this phenomena is  used for the early detection of flu outbreaks among college students. In~\cite{GFP_disaster}, it is utilized to efficiently sample early-warning sensors during catastrophic events such as hurricanes. 

In addition to degree, the same paradox has been observed about other individual attributes (called the \emph{generalized friendship paradox}~\cite{GFP_nature}, or GFP). For example, in~\cite{Hodas_twitter} it has been observed that on Twitter, for most people, their friends share, on average,  more viral content and also tweet more. In~\cite{GFP_nature}, it has been observed that in scientific collaboration networks, one's co-authors have, on average,  more citations, more publications and more co-authors.

In this paper, we   consider a network growth model which is a  generalization of the preferential attachment scheme~\cite{BA_1}.
In our model,  nodes are endowed with `qualities' (ak.a.~`fitness' or `attractiveness' in the literature~\cite{Bianconi,fit_1,fit_2,fit_3}). Qualities are discrete positive numbers drawn from a given distribution $\rho(\theta)$ and assigned to a node upon its birth (remaining the same thenafter).  We assume that the  probability that   node $x$ with degree $k_x$ and quality $\theta_x$ receives a link from subsequent nodes is proportional to $k_x+\theta_x$.\footnote{Note that for example in~\cite{Bianconi}, the  attachment  probability is proportional to the product of degree and quality. This model however, has not be solved in closed form. Also, it assigns zero link reception probability to nodes with degree zero.} We obtain two statistical measures of this model: one is the degree-quality joint distribution, which is the fraction of nodes that have degree $k$ and quality $\theta$ in the steady state. The second quantity is the nearest-neighbor distribution of quality and degree: it gives the fraction of nodes with degree $\ell$ and quality $\phi$ that are connected to a node with degree $k$ and quality $\theta$.  Equipped with these distributions, we can quantify the paradox and study how it depends on the underlying quality distribution $\rho(\theta)$. To our knowledge, no similar theoretical result is available in the literature for any network growth model (either purely preferential~\cite{BA_1}, or fitness-based~\cite{fit_1,fit_2,fit_3}). 

 We show that employing the above scheme as the attachment mechanism renders the occurrence of the GFP contingent upon the underlying distribution of node qualities. 
 We then employ measures defined in the literature for assessing the GFP on the network level, and we investigate the dependence of these measures on the model parameters and the quality distribution.  We demonstrate that, in the proposed model, the network exhibits a quality paradox at the network level for any quality distribution.  We contend that this is indicative of a positive correlation between degree and quality; i.e., those with higher qualities are more likely to have higher degrees, and vice versa.

\section{Model, Notation and Terminology}
In the growth model considered in this paper, nodes are added successively to the network. 
The initial network has $N(0)$ nodes and $L(0)$ links. At each time step, one new node is added to the network.  We assume that each node has an intrinsic quality, which is drawn from a given distribution $\rho(\theta)$. The quality is assigned to each new incoming node upon  birth, and will remain the same thenafter. The mean of the distribution $\rho(\theta)$ is denoted by $\mu$. 
 A  node of degree $k$ and quality $\theta$ is also referred to as   \emph{a $(k,\theta)$ node} throughout.

Each new incoming node attaches to $\beta \leq N(0)$ existing nodes in the network. 
We consider the simplest additive model that incorporates both degree (popularity) and quality in the dynamics of connection formation: the probability that an existing node with degree $k$ and quality $\theta$ receives a link from the new node  is  proportional  to ${k+\theta}$.  This means that, for example, a paper that is new and has very few citations can compensate for its small degree with having a high quality. Or in the social context, a newcomer who does not have many friends in the new social milieu but is gregarious and sociable can elevate the chances of making new friends. The new node is called the \emph{child} of the existing nodes that it connects to, and they are called its \emph{parents}. By \emph{a  {$(\ell,\phi)$-$(k,\theta)$} child-parent pair}, we mean a  node  with  degree $\ell$ and quality $\phi$ that is connected to a parent node of degree $k$ and quality $\theta$.

The probability that an existing node $x$ receives a new link is $\frac{k_x+\theta_x}{A}$, where the normalization factor $A$ is given by ${\sum_x  (k_x+\theta_x)}$. The sum over all node degrees at time $t$, which equals twice the number of links at time $t$, is equal to $2[L(0)+\beta t]$. For long times, the sum over the quality values of all the nodes will converge to the mean of the quality distribution times the number of nodes, that is, we can replace ${\sum_x \theta_x}$ by $[N(0)+t]\mu$. So at time $t$, the probability that node $x$ receives a link equals $\frac{k_x+\theta_x}{2L(0)+N(0)+(2\beta+\mu)t}$. 

Throughout the present paper, the steady-state joint distribution of quality and degree is denoted by $P(k,\theta)$. The expected number of nodes with degree $k$ and quality $\theta$ at time $t$ is denoted by $N_t(k,\theta)$. We denote by $N_t(k,\theta,\ell,\phi)$ the expected number of 
{$(\ell,\phi)$-$(k,\theta)$} child-parent pairs.

\section{Degree-quality Joint Distribution}
We seek the steady-state fraction of nodes who have degree $k$ and quality $\theta$.  
In Appendix~\ref{app:sol_1} we derive the following expression for this quantity:
\all{
P(k,\theta)
 = 
 \rho(\theta) \left( 2+\frac{\mu}{\beta}\right) 
  \frac{\Gamma(k+\theta)}{\Gamma(\beta+\theta)}
 \frac{\Gamma \left(\beta+\theta+2+\fracc{\mu}{\beta} \right) }
{\Gamma \left( k+\theta+3+\fracc{\mu}{\beta} \right) }  
  u(k-\beta).
}{Pkth_fin}

Note that in the special case of a single permitted value for the quality (that is, when $\rho(\theta)=\delta[\theta-\theta_0]$) this model reduces to the shifted-linear preferential attachment model analyzed, for example, in~\cite{ME_EPJB}. The solution in this special case simplifies to 
\al{
P_{sh}(k)
 = 
   \left(2+\frac{\theta_0}{\beta}\right)     
 \frac{\Gamma(k+\theta_0)}{\Gamma(\beta+\theta_0)}
 \frac{\Gamma (\beta+2+\theta_0+\frac{\theta_0}{\beta} ) }
{\Gamma ( k+3+\theta_0+\frac{\theta_0}{\beta} ) }
.}
%
%
This coincides with the degree distribution of shifted-linear kernels given in~\cite{dorog_rate_1} and~\cite[Equation~D.9]{ME_EPJB}. Furthermore, when $\rho(0)=1$, all nodes will have zero quality and attachments will be purely degree-proportional, synonymous with the conventional preferential-attachment model proposed initially in~\cite{BA_1}.  For the special case of $\theta=\mu=0$ we obtain
 \al{  
 P_{BA}(k)  =     \frac{ 2 \beta (\beta+1 )}{k (k+1 )(k+2 )}  
 . }
This is equal to the degree distribution of the conventional BA network (see, e.g.,~\cite{dorog_rate_1,redner}). 

Let us also examine the behavior of~\eqref{Pkth_fin} in the limit of large $k$. In this regime, we can use the asymptotic approximation that for large values of $x$, the function $\Gamma(x) \approx x^{x-\frac{1}{2}} \exp(-x)$. Then we replace $\frac{\Gamma(k+\theta)}{\Gamma(k+ \theta+3+\frac{\mu}{\beta})}$  with   $k^{-3-\frac{\mu}{\beta}}$, independent of $\theta$. Therefore, the steady-state joint degree-quality distribution $P(k,\theta)$ is proportional to $k^{-3-\frac{\mu}{\beta}}$. Marginalizing out $\theta$ to recover the degree distribution, we obtain the well-known power law, $P(k)=k^{-3-\frac{\mu}{\beta}}$. 

%
%

\section{Nearest-Neighbor Quality-Degree Distribution}

To quantify how qualities and degrees of adjacent nodes correlate, we need to go beyond the quality-degree distribution obtained in the previous section. The closed-form expression for the nearest-neigbor correlations under the preferential attachment model is derived in \cite{ME_EPJB}; that work only considers degrees and does not address qualities. We would like to quantify the conditional distribution $P(\ell, \phi | k, \theta)$, the fraction of neighbours of a given node with degree $k$ and quality $\theta$ that have degree $\ell$ and quality $\phi$. We refer to this as the \emph{nearest-neighbor quality-degree distribution} (NNQDD).

In Appendix~\ref{app:sol_2} we study the rate equation describing how the distribution $P(\ell, \phi | k, \theta)$ evolves as nodes are added to the network. This gives rise to a system of difference equations which we solve to obtain that, in the steady-state,
\all{
  &P(\ell,\phi|k,\theta)=
\fracc{\rho(\phi)  }{k}
  \frac{ \Gamma \left(k+\theta+3+\fracc{\mu}{\beta} \right)}{\Gamma\left( k+\theta+3+\frac{\mu}{\beta}+\ell+\phi  \right)  }
\frac{   (\ell-1+\phi)! }{  (\beta-1+\phi)!  }  
\Gamma \left(\beta+2+\phi+\fracc{\mu}{\beta} \right)\times
\nonumber \\ 
&
\resizebox {\linewidth}{!}{$
\left[ \DD
 \sum_{j=\beta+1}^{k} \frac{\Gamma \left(j+\theta+2+\fracc{\mu}{\beta}+\beta+\phi \right) \DD \CC{k-j+\ell-\beta}{\ell-\beta}
}{\Gamma \left( j+\theta+2+\fracc{\mu}{\beta} \right)\Gamma \left(\beta+2+\phi+\fracc{\mu}{\beta} \right) } 
 +  \DD \sum_{j=\beta+1}^{\ell} \frac{\Gamma \left(j+\theta+2+\fracc{\mu}{\beta}+\beta+\phi \right)\DD \CC{\ell-j+k-\beta}{k-\beta}
}{ \Gamma \left( j+\phi+2+\fracc{\mu}{\beta} \right)\Gamma \left(\beta+2+\theta+\fracc{\mu}{\beta} \right) } 
\right].$}}{NNQDD}
%

%
%
%
%
%
In order to obtain the nearest-neighbor quality distribution $P(\phi|\theta)$, one needs to perform the calculations ${ P(\phi|\theta)=    \sum_{\ell} \sum_{k} P(k) P(\ell,\phi|k,\theta) }$, which requires knowledge of $P(k)$. In turn we have ${P(k)=\sum_{\theta} P(k,\theta)}$, which according to~\eqref{Pkth_fin},  yields different sums for different quality distributions $\rho(\theta)$.

\section{Quantifying the Friendship and Generalized Friendship Paradoxes}

As discussed in Section~\ref{sec:intro},  GFP  refers to an average alter superiority in arbitrary aspects (e.g., number of citations, exposure to viral online content). In this paper, we use the `quality' dimension that is incorporated in the model as the subject of the GFP.  
 Our objective is to compare the degrees and qualities of nodes with their neighbors. We say that a node experiences the friendship paradox if the degree of that node is less than the average of the degrees of its neighbors. Similarly, we say that a node experiences the quality paradox if the quality of the node is less than the average of the qualities of its neighbors.

The above-mentioned definitions characterize individual-level paradoxes. Our primary interest is to what fraction of nodes experience the friendship and quality paradoxes. To this end, we compare the  average degree of the nodes with the average degree of the neighbors of all nodes (and similarly for quality). Comparing these two average values yields a macro measure for the system, indicating whether it exhibits paradoxes on average. We call these as the \emph{network-level friendship paradox}  and \emph{network-level quality paradox}.

Our measure of the network-level quality paradox is defined as $\textnormal {NQP} = \frac{  \sum_i k_i \theta_i}{  \sum_i k_i} - \frac{1}{N}   \sum_i \theta_i$. 
The summations are performed over all nodes in the network. Note that the numerator of the first sum is actually the sum of the qualities of the neighbors of all nodes. Node $i$ is repeated $k_i$ times in this sum, once for each of its neighbors. Focusing on the limit as ${t \rightarrow \infty}$, we can use  the law   of large numbers and  express the NQP  as follows
\all{
\textnormal {NQP} = \fracc{  \sum_{k,\theta} k \theta P(k,\theta)}{  \sum_{k,\theta} k P(k,\theta)} - \mu
.}{NQP_def_2}
The greater NQP becomes, the more strongly the paradox holds. Negative NQP is indicative of the absence of a quality paradox at the network level. 

Undertaking similar steps to above, we can measure the network-level friendship paradox via
\all{
\textnormal{NFP} = \fracc{\ave{k^2}}{\ave{k}}-\ave{k} = \fracc{\ave{k^2}-\ave{k}^2}{\ave{k}}
.}{NFP_def_1}

Note that the numerator is the variance of the degree distribution, so it is positive. The denominator  is the average degree and is also positive. So  the  NFP  is always positive, which means that by this definition:  \emph{any network exhibits the friendship paradox at the network level. } So the task of the present paper with regard to the NFP is to investigate its magnitude, i.e., to measure how strongly the paradox holds. For example, in the conventional Barabasi-Albert scale-free model, where the degree variance diverges, the NFP also diverges, which is a result of the presence of macro hubs. 


\section{Results  and Discussion }
To study the NFP and the NQP in concrete settings, we  confine ourselves  to two quality distributions $\rho(\theta)$ for illustrative purposes. We consider a finite support for $\theta$, so that ${0\leq \theta \leq \theta_{\textnormal{max}}}$. For each distribution, we are going to consider four different values $\beta$, and four different values of $\theta_{\textnormal{max}}$. 

\begin{figure}[!Ht]
        \centering
      \begin{subfigure}[b]{.48  \textwidth}
              \includegraphics[width=  \textwidth,height=4cm]{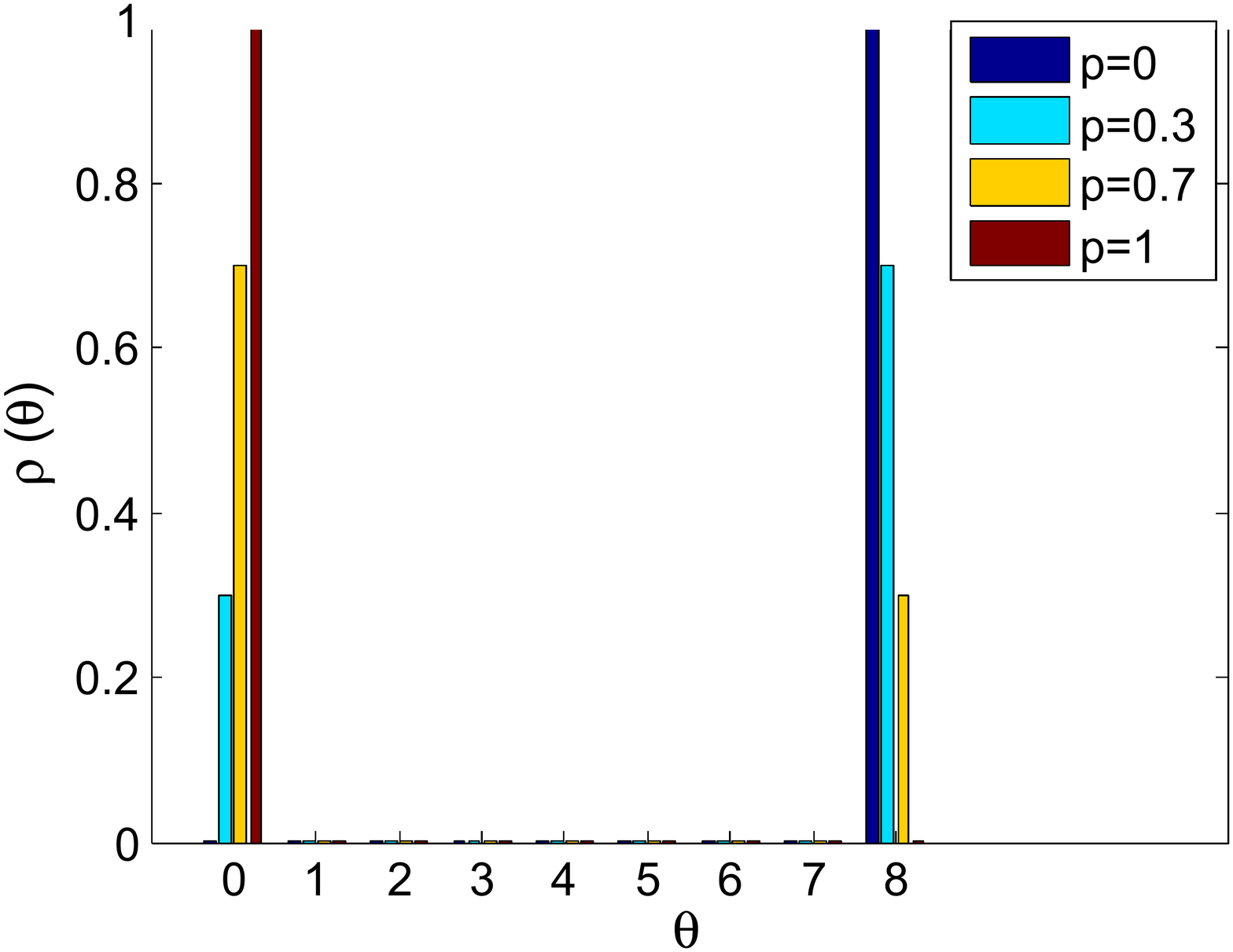}
                \caption{
    Bernoulli distribution with  ${p=0,0.3,0.7,0.1}$. The cases of ${p=0}$ and ${p=1}$ correspond to conventional Barabasi-Albert and shifted-linear preferential attachment networks, respectively. 
}
                \label{distribution_extreme}
        \end{subfigure}%
     ~~~   \begin{subfigure}[b]{0.48\textwidth}
                \includegraphics[width=  \textwidth,height=4cm]{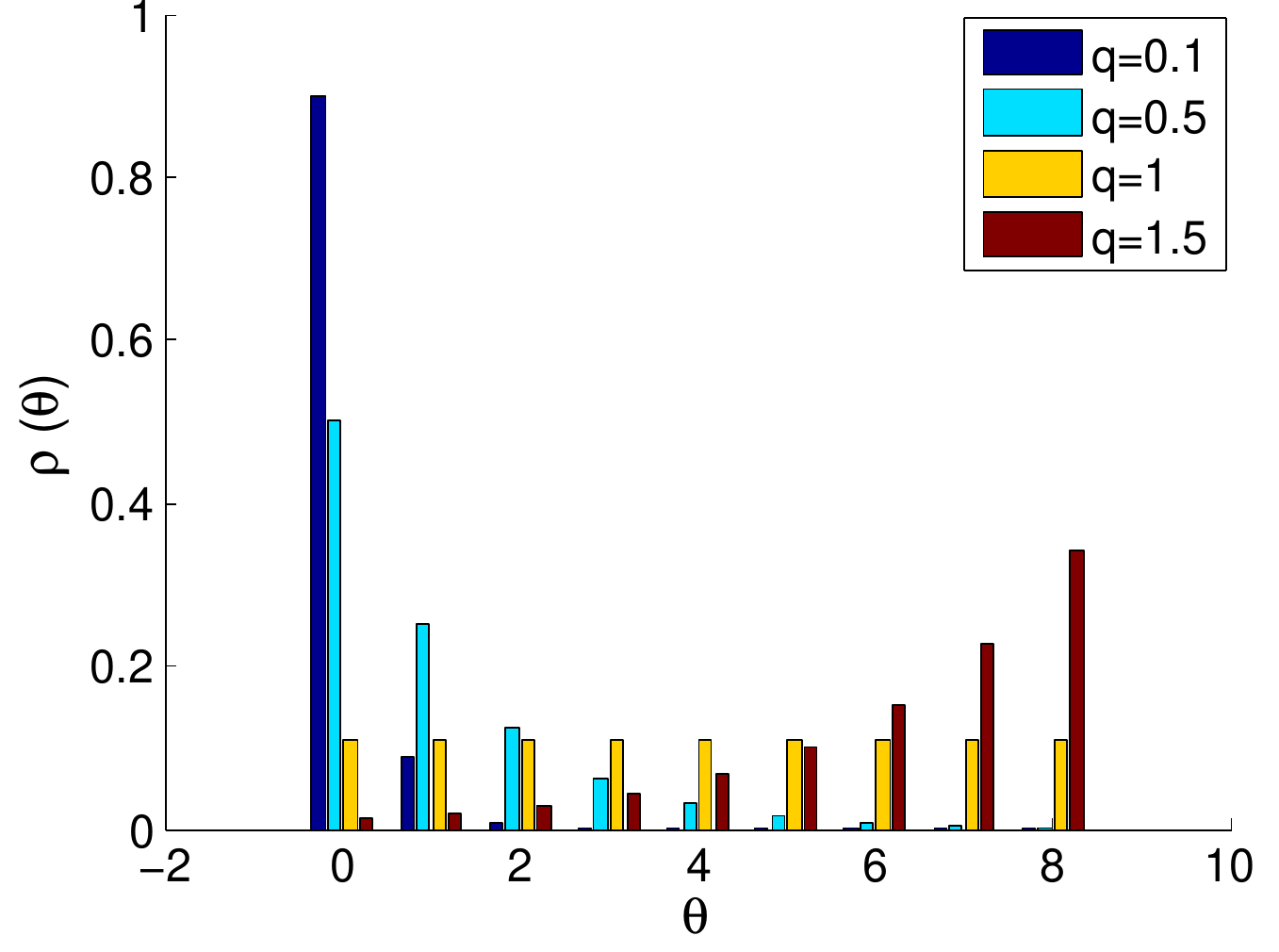}
                \caption{
Exponential distribution for decay factor ${q=0.1,0.5,1,1.5}$. The special case of ${q=1}$ corresponds to a uniform distribution supported in the interval ${0\leq \theta \leq \thmax}$. 
}
                \label{distribution_decay}
        \end{subfigure}%
\caption{Examples of the quality distributions used in this paper with ${\thmax=8}$. Four instances of each type is depicted.}
\label{distributions}
\end{figure}

The first distribution we consider is the Bernoulli case, where nodes can either have quality zero or quality $\thmax$. The probability of quality zero is $p$ and the probability of quality $\thmax$ is  ${1-p}$, where ${0\leq p \leq 1}$. The second distribution we consider is the discrete exponential distribution with decay factor $q$. The probability that the quality is $\theta$ is proportional to $q^{\theta}$. Note that in the case of $q=1$, one recovers a uniform distribution as a special case. We consider both $q<1$ and $q>1$, yielding  decreasing and increasing distributions in $\theta$, respectively. These distributions are depicted in Figure~\ref{distributions}. 

\begin{figure}[!Ht]
        \centering
      \begin{subfigure}[b]{.5 \textwidth}
              \includegraphics[width=\textwidth,height=4.25cm]{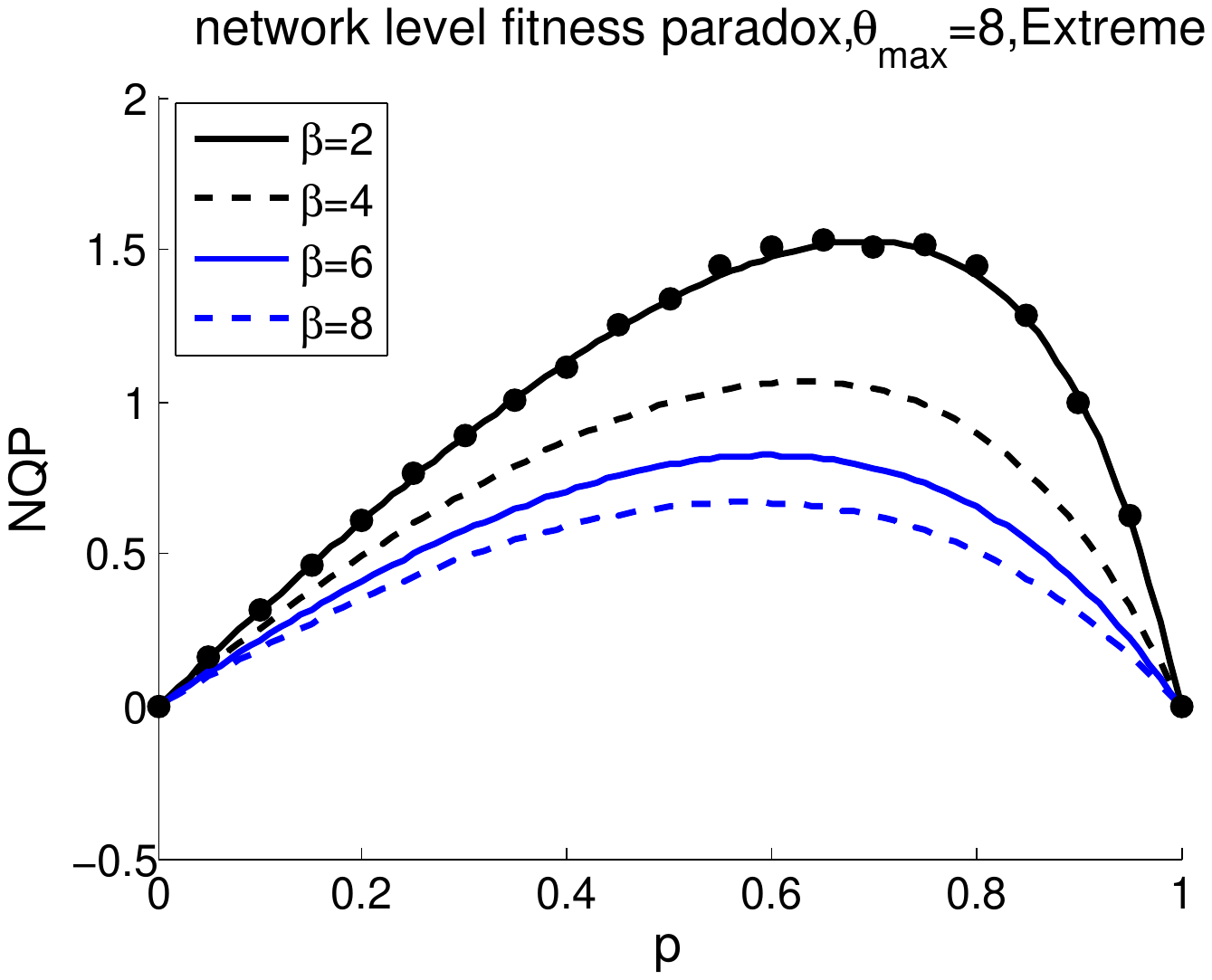}
                \caption{
$\thmax=8$
}
                \label{NQP_fixed_theta}
        \end{subfigure}%
        \begin{subfigure}[b]{0.5\textwidth}
                \includegraphics[width=\textwidth,height=4.25cm]{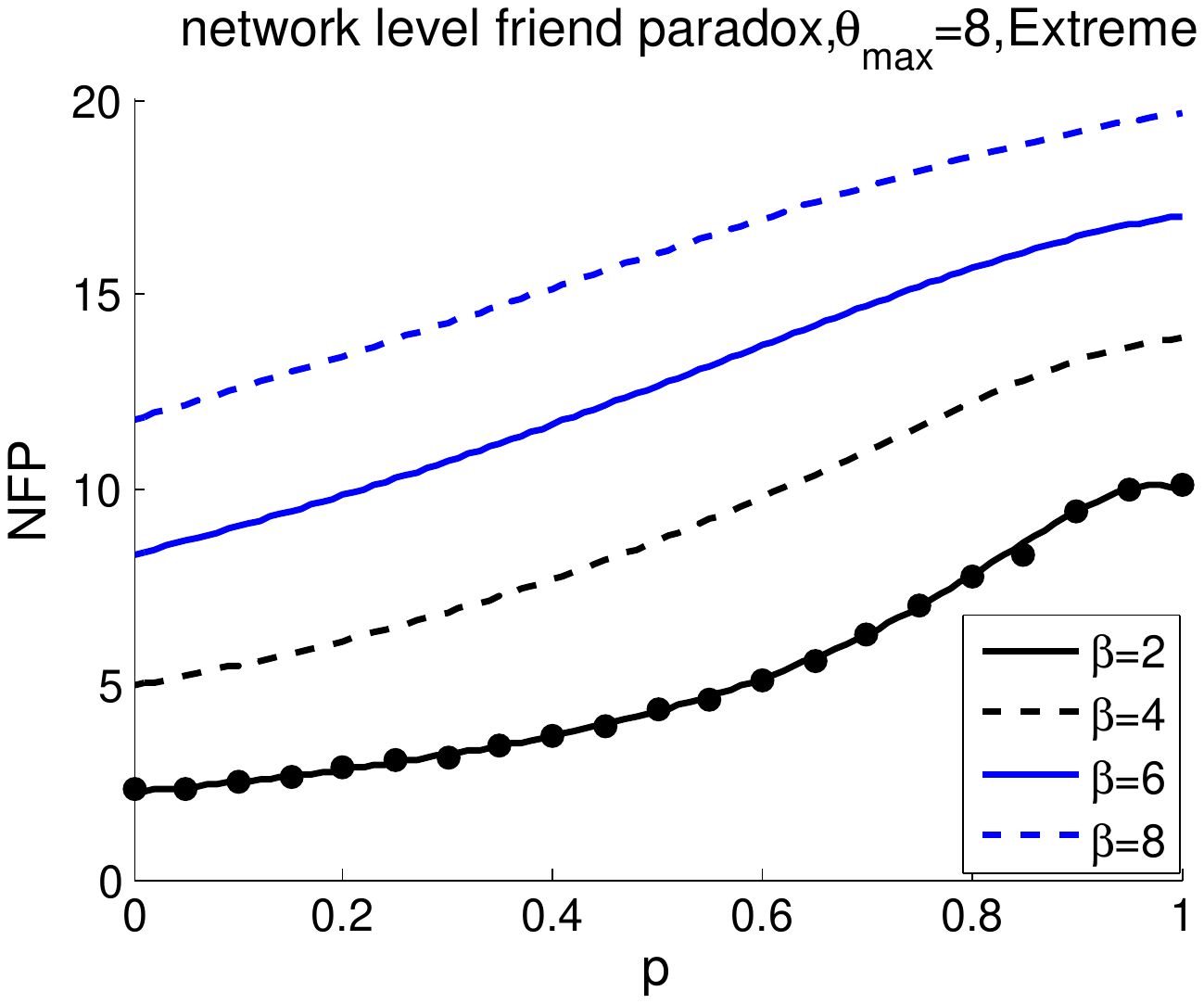}
                \caption{$\thmax=8$}
                \label{NFP_fixed_theta}
        \end{subfigure}%
\\
 \begin{subfigure}[b]{0.5\textwidth}
                \includegraphics[width=\textwidth,height=4.25cm]{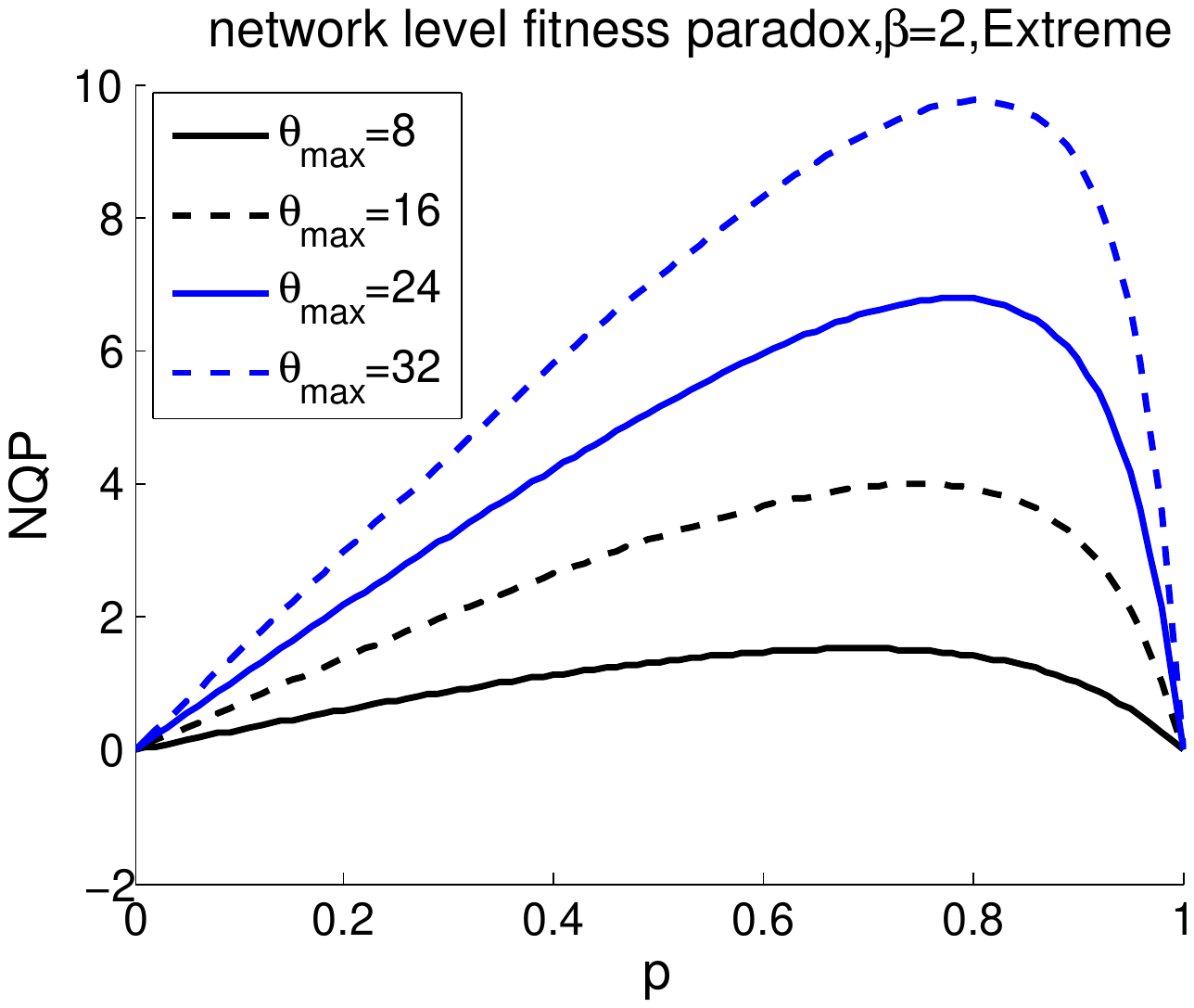}
                \caption{$\beta=2$}
                \label{NQP_fixed_beta}
        \end{subfigure}%
        \begin{subfigure}[b]{0.5\textwidth}
                \includegraphics[width= \textwidth,height=4.25cm]{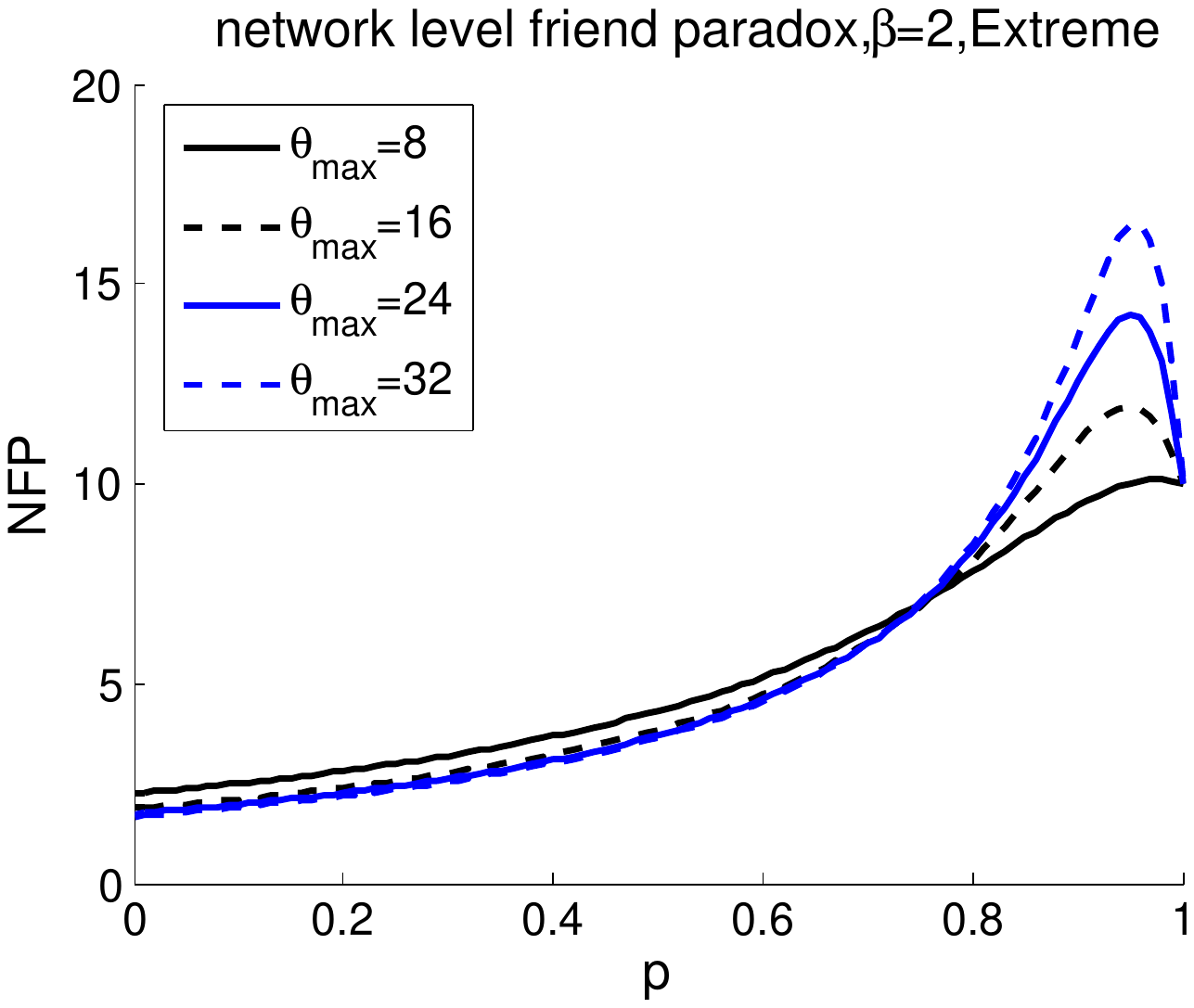}
                \caption{$\beta=2 $}
                \label{NFP_fixed_beta}
        \end{subfigure}
         \caption{
          Network level friendship and quality paradox for Bernoulli quality distribution. The markers in Figures (a) and  (b) represent simulation results, and the solid curves are the theoretical expression. The depicted results are averaged over 100 Monte Carlo trials.
          }
\label{bernoulli}
\end{figure}

The results for the Bernoulli quality distribution are depicted in Figure~\ref{bernoulli}. As depicted in Figure~\ref{NQP_fixed_theta}, for a fixed $\thmax$, the NQP decreases as $\beta$ (the initial degree of nodes)  increases. Also, it is observable that   the sensitivity of the NQP to the variations of the quality distribution diminishes for larger values of $\beta$.

As illustrated in Figure~\ref{NFP_fixed_theta}, the NFP increases as $\beta$ (the initial degree  of nodes)  increases.  Hence, according to~\eqref{NFP_def_1}   \emph{the variance of the degree distribution grows faster than the mean degree, as $\beta$ increases.} On the other hand, for a given $\beta$, increasing $\thmax$ (which is tantamount to increasing $\mu$), increases the NQP. This means that according to~\eqref{NQP_def_2} \emph{as $\thmax$ increases,  the mean of the qualities of the neighbors increases faster than the mean of the qualities of the nodes}.

Figure~\ref{NQP_fixed_beta} pertains to this case. Observe that as $\thmax$ increases, the NQP becomes more sensitive to the distribution of qualities.  Finally,  Figure~\ref{NFP_fixed_beta} represents the  NFP  for a fixed $\beta$ and different values of $\thmax$. From Figures~\ref{NQP_fixed_theta},~\ref{NFP_fixed_theta},~\ref{NQP_fixed_beta} and~\ref{NFP_fixed_beta}, a general observable pattern is that as $p$ increases, the NFP increases (monotonically for almost all values of $p$), whereas   the NQP  is concave and unimodal (it increases at first, achieves maximum, and then decreases).  

\begin{figure}[!Ht]
        \centering
      \begin{subfigure}[b]{.5 \textwidth}
              \includegraphics[width= \textwidth,height=4.25cm]{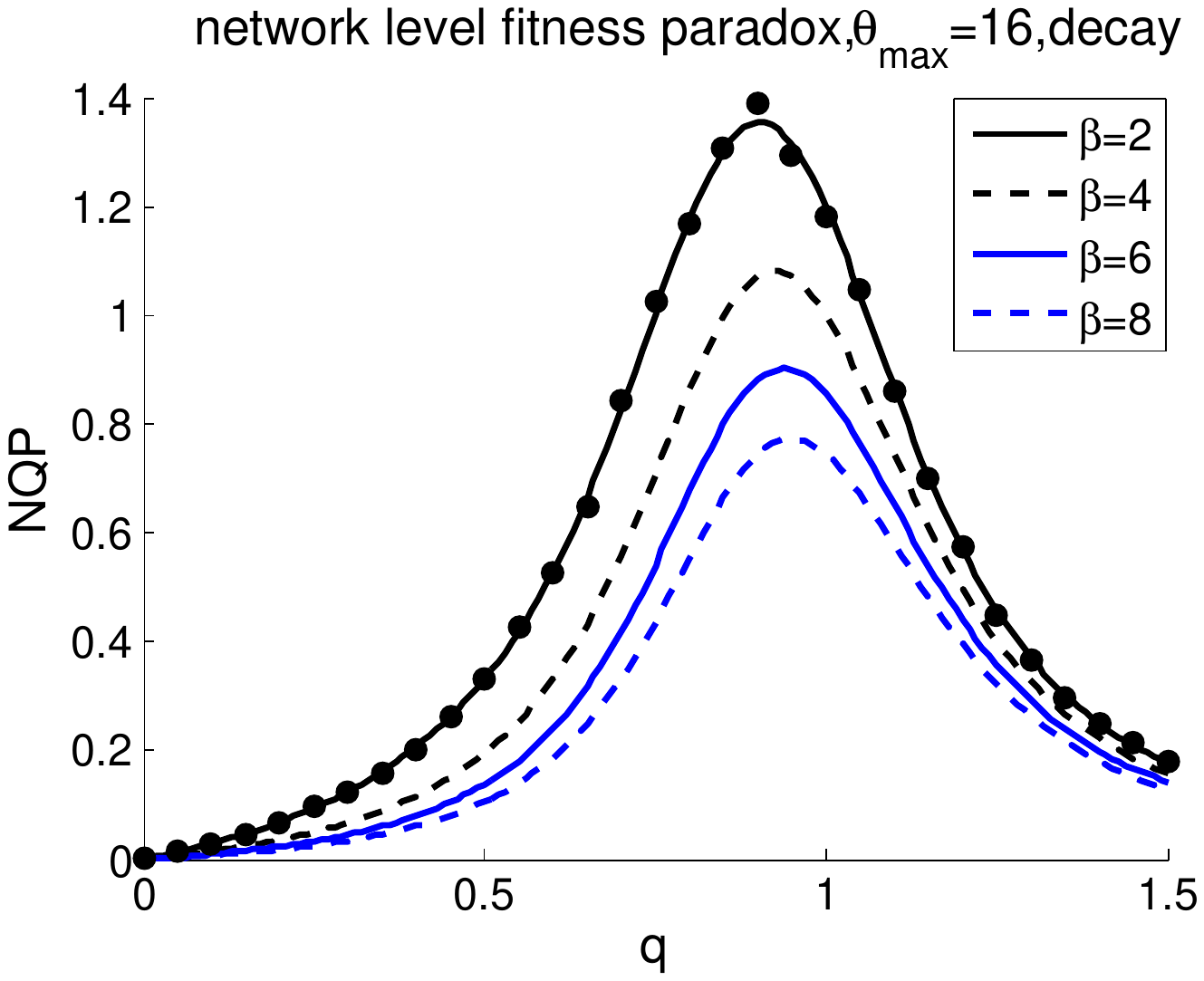}
                \caption{
$\theta=16$
}
                \label{NQP_fixed_theta_decay}
        \end{subfigure}%
        \begin{subfigure}[b]{0.5\textwidth}
                \includegraphics[width= \textwidth,height=4.25cm]{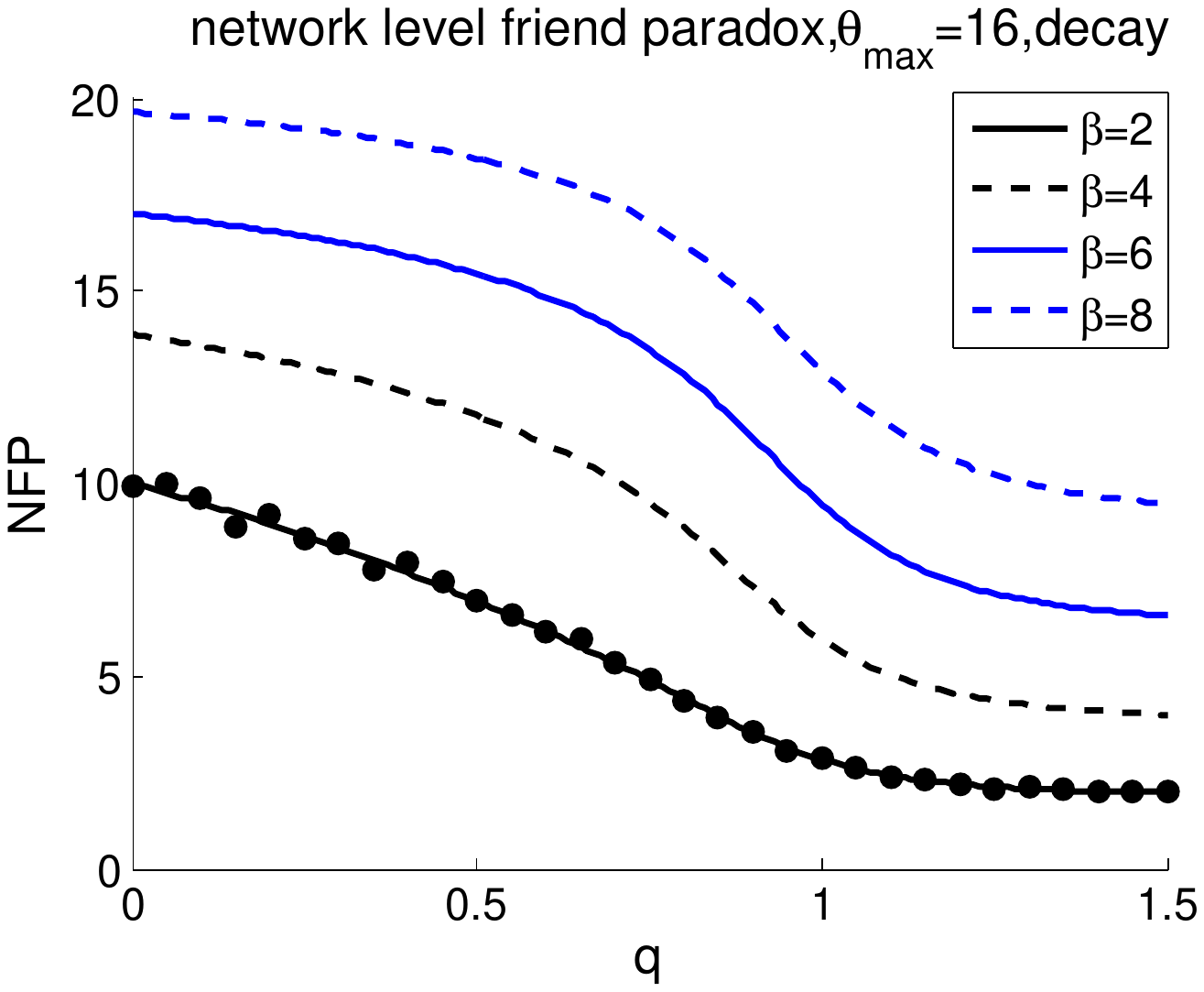}
                \caption{$\theta=16$}
                \label{NFP_fixed_theta_decay}
        \end{subfigure}%
\\
 \begin{subfigure}[b]{0.5\textwidth}
                \includegraphics[width= \textwidth,height=4.25cm]{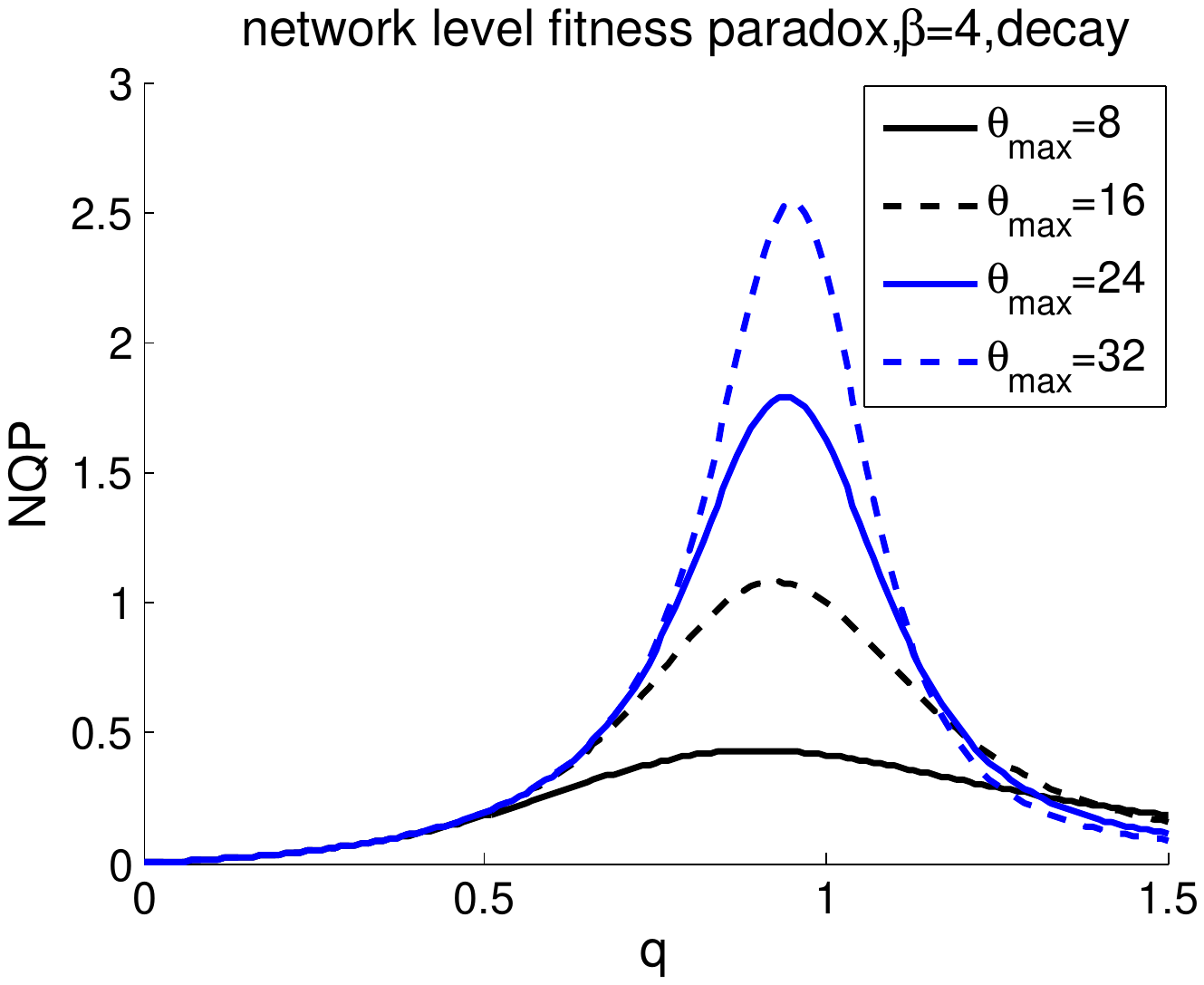}
                \caption{$\beta=4$}
                \label{NQP_fixed_beta_decay}
        \end{subfigure}%
        \begin{subfigure}[b]{0.5\textwidth}
                \includegraphics[width=  \textwidth,height=4.25cm]{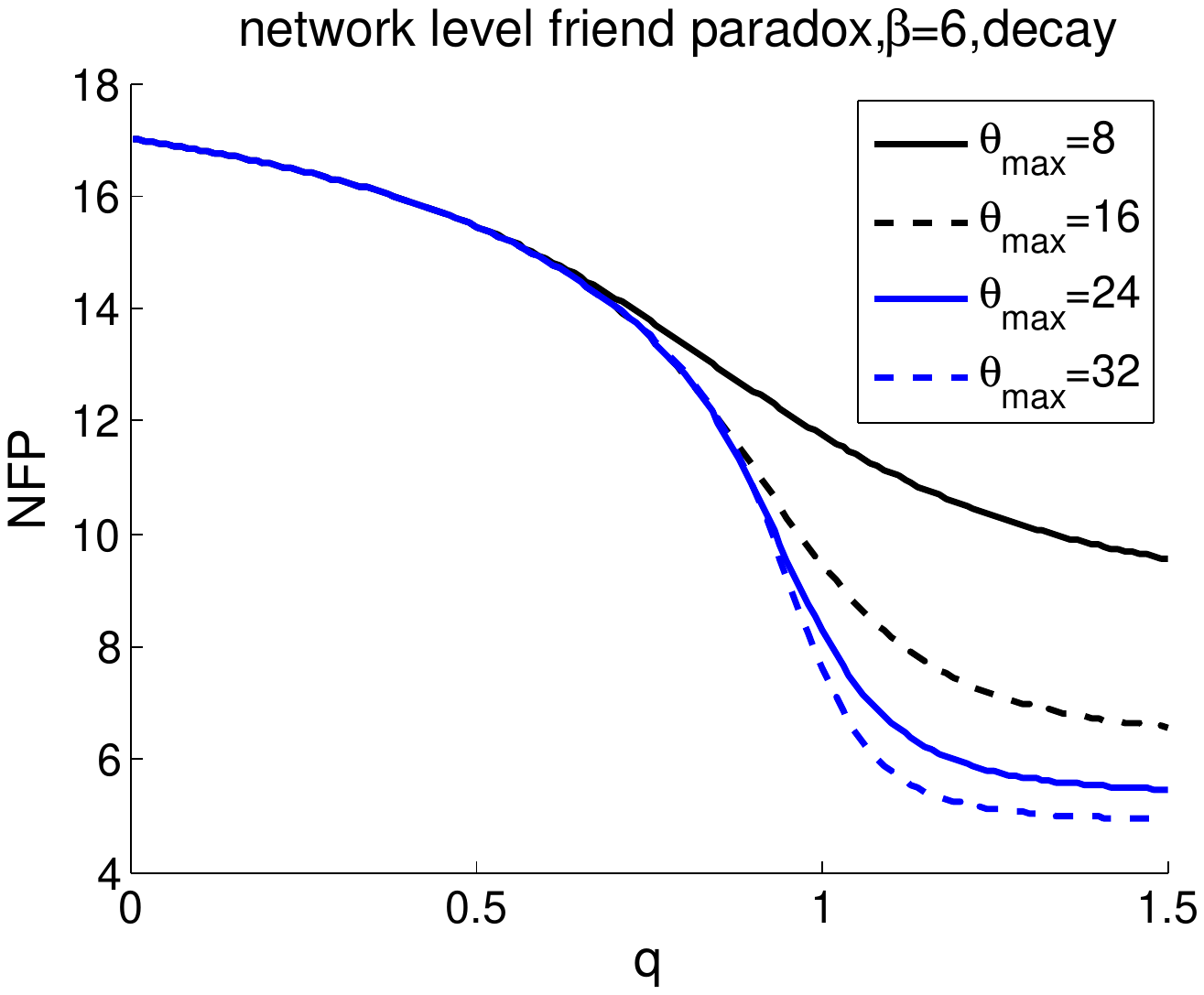}
                \caption{$\beta=6$}
                \label{NFP_fixed_beta_decay}
        \end{subfigure}
         \caption{
                  Network level friendship and quality paradox for  exponential  quality distribution. The markers in Figures (a) and  (b) represent simulation results and the solid curves are from the theoretical expressions. The depicted results are averaged over 100 Monte Carlo trials.  
          }
\label{PLK_second_term}
\end{figure}

Now we focus on the exponential quality distribution with the decay factor denoted by $q$. As depicted in Figure~\ref{NQP_fixed_theta_decay}, for a given $\thmax$, the NQP decreases as $\beta$ increases. Also, it is observed that as $\beta$ increases,  the sensitivity of the NQP to the quality distribution diminishes. These are both similar to the results of the Bernoulli distribution. 
As can be seen in  Figure~\ref{NFP_fixed_theta_decay}, the NFP increases as $\beta$   increases.  So similar to the Bernoulli case,  the variance of the degree distribution grows faster than the mean degree, as $\beta$ increases.

From  Figure~\ref{NQP_fixed_beta_decay} we observe that  for a  fixed $\beta$, increasing $\thmax$ increases the NQP.    We    observe that as $\thmax$ increases, NQP becomes more sensitive to the  changes in the decay factor.  Finally,  Figure~\ref{NFP_fixed_beta_decay} represents the  NFP  for a fixed $\beta$ and different values of $\thmax$.  We observe  that increasing $\thmax$ increases the  NFP for positive decays. Also, for very small decay factors (which generate right-skewed distributions that are highly unequal), changing $\thmax$ has scant  effect on  the NFP. This is reasonable because when the decay factor is  small, all large values of $\theta$ have  small chances  of occurrence. Consequently,  changing $\thmax$ minimally changes the shape of the distribution for small decay factors. 

A trend  is discernible from   Figures~\ref{NQP_fixed_theta_decay},~\ref{NFP_fixed_theta_decay},~\ref{NQP_fixed_beta_decay} and~\ref{NFP_fixed_beta_decay}:  as $q$ increases, the NFP decreases (monotonically for all values of $q$), whereas   NQP  is concave and  increases up to  a point around ${q=1}$, and then decreases.  Since $q=1$ yields a uniform distribution, we can qualitatively conclude that the probability of the network-level quality paradox is higher when qualities are heterogeneous, as compared to when qualities are similar.

Finally, to verify our results, we run Monte Carlo simulations to synthesize networks that grow under the prescribed quality-based preferential attachment mechanism, and then calculate the desired quantities by averaging over nodes in the synthesized network. Due to computational limitations, we restrict this validation to the case where $\beta = 2$ and $\theta_{\max} = 8$ for the Bernoulli quality distribution and the case where $\beta = 2$ and $\theta_{\max} = 16$ for the  exponential quality distribution. These results are shown in Figures~\ref{NQP_fixed_theta} ,~\ref{NFP_fixed_theta},~\ref{NQP_fixed_theta_decay} and~\ref{NFP_fixed_theta_decay}. The markers show the results of simulations, averaging over 100 Monte Carlo trials, and the solid curves correspond to our theoretical expressions.

We have tested the results on various other quality distributions and observed similar results; these additional simulations not reported here due to space limitations. In general, we observe that for a fixed $\thmax$, increasing $\beta$ increases the NFP and decreases the NQP regardless of the quality distribution. Also, for a fixed $\beta$, increasing  $\thmax$ increases the NQP and decreases the NFP.

Note that in all cases the NQP is nonnegative. This has roots  in the correlation between degree and quality of single nodes (intra-node correlation, rather than inter-node correlation). Let us denote the correlation between degree and quality for a node by $\rho_{k \theta}$, which is the Pearson correlation  coefficient obtained from the joint distribution $P(k,\theta)$. From~\eqref{NQP_def_2}, we have:
\all{
\textnormal {NQP} &= \fracc{  \sum_{k,\theta} k \theta P(k,\theta)}{  \sum_{k,\theta} k P(k,\theta)} - \mu
=\fracc{\sum_{k,\theta} k \theta P(k,\theta) - \mu  \sum_{k,\theta} k P(k,\theta)}{ \sum_{k,\theta} k P(k,\theta)}
\nonumber \\ &
=\fracc{\sum_{k,\theta} k \theta P(k,\theta) - \mu \ave{k}}{ \ave{k}} 
= \fracc{\rho_{k \theta} \sigma_k \sigma_{\theta}}{\ave{k}}
.}{NQP_corr}

This implies that the sign of NQP is the same as the sign of $\rho_{k \theta}$ (since $\sigma_k,\sigma_{\theta}$ and $\ave{k}$ are  nonnegative). The observation that  NQP is always nonegative indicates that $\rho_{k \theta}$ is also always nonegative. We  conclude that \emph{the quality-dependent preferential attachment model generates networks in which degree and quality of a node are always positively correlated.} This is what we intuitively expect the model to exhibit; increasing quality increases degree. For example, in citation networks, papers with higher qualities receive more citations. Conversely, a paper with many citations is more likely to have a high quality. In the case of friendship networks, a person that is more sociable ends up with more friends than an anti-social person, and conversely, a popular person is more likely to be friendly than an isolated person. 

We also observe that in all cases, $\mu$ (equivalently, $\thmax$) and $\beta$ have opposite effects on both the NFP and the NQP. That is,  the effect of increasing $\beta$ is akin to that of decreasing $\mu$, and vice versa. We observed similar trends for other quality distributions; these results are omitted here due to space limitations. What causes this disparity is the following:  as  can  be seen in~\eqref{Pkth_fin} and~\eqref{NNQDD}, $\mu$ only appears in the distributions in the form of $ \frac{\mu}{\beta}$. Thus increasing $\mu$ and decreasing $\beta$ have the same effect on this variable, and consequently, on the distribution.

\section{Summary and Future Work}
The aim of the present paper was to put in crisp theoretical focus the seemingly prevalent phenomena  of the friendship paradox and the generalized friendship paradox. We proposed a network growth model that incorporates quality. In this model, the probability that a node receives a link increases with both its degree and quality. We analysed the model theoretically in the steady-state (large size limit), and found two theoretical quantities that characterize the interrelation between quality and degree. The first quantity is $P(k,\theta)$, which is the joint degree-quality distribution, and equals the fraction of nodes who have degree $k$ and quality $\theta$. The second quantity characterizes nearest-neighbor correlations, and is the nearest-neighbor  quality-degree distribution, denoted by ${P(\ell,\phi|k,\theta)}$.

We then defined two network-level measures for the quality and friendship paradoxes and computed them for two  particular examples  of quality distributions. 
 We observed that for a fixed $\thmax$,  increasing $\beta$ increases the NFP and decreases the NQP  regardless of the quality distribution. We also observed that  for a fixed $\beta$, increasing  $\thmax$  increases the NQP and decreases the NFP. We also observed that $\mu$ and $\beta$ have opposite effects on the NFP and also on the NQP. We also tested these   results on various other quality distributions, and they proved robust; the  effects of $\beta$ and $\mu$ on paradoxes are opposite regardless  of the quality distribution. 

There are many interesting extensions of this work to pursue. In addition to the network-level paradox, we can also study the individual-level paradox, which would require the utilization of the NNQDD to compare the degrees and qualities of nodes with those of their neighbors. 
 The individual-level paradox has empirical implications which enable us to assess the quality distribution of real networks.

\newpage
\appendix
\section{Obtaining the Joint Distribution $P(k,\theta)$}\label{app:sol_1}

We seek the fraction of nodes who have degree $k$ and have quality $\theta$.  We begin by writing the rate equation which quantifies the temporal evolution of $N_t(k,\theta)$. Suppose that a node with quality  $\theta$ and degree $k-1$ at time $t-1$, receives a link from the new incoming node. Consequently, its degree will become $k$ and $N_t(k,\theta)$ increments. Conversely, if  a node with quality  $\theta$ and degree $k$ at time $t-1$, receives a link from the new incoming node, $N_t(k,\theta)$ decrements. Finally, each new incoming node increments $N_t(\beta,\theta)$ with probability $\rho(\theta)$. The rate equation thus reads
\all{
N_{t+1}(k,\theta)-N_t(k,\theta) &= 
\fracc{\beta(k-1+\ttt) N_t(k-1,\theta)}{ \lam+(2\beta+\mu)   t}
 \nonumber \\ &
 - \fracc{\beta(k +\ttt) N_t(k ,\theta)}{\lam+(2\beta+\mu)  t}
+ \rho(\ttt) \delta_{k,\beta}
.}{Nkdot_1_0}

Replacing $N_t(k,\theta)$ by $[N(0)+t] P_t(k,\theta)$, this can be   expressed in terms of $P_t(k,\theta)$ as follows:
\all{
&\big[N(0)+t \big] \big[ P_{t+1}(k,\theta)-P_t(k,\theta) \big] + P_{t+1}(k,\theta)=
\nonumber \\  &
\resizebox{0.93\linewidth}{!}{$\fracc{\beta(k-1+\ttt)  [N(0)+t] P_t(k-1,\theta)}{\lam+(2\beta+\mu)   t}
  - \fracc{\beta(k +\ttt) [N(0)+t] P_t(k ,\theta)}{\lam+(2\beta+\mu)  t}
+ \rho(\ttt) \delta_{k,\beta}
.$}}{Pkdot_1_0}
In the limit as $t \rightarrow \infty$, the transients vanish.  So,   we   drop the $t$ in the arguments and rewrite~\eqref{Pkdot_1_0} as:
\all{
 P( k,\theta)
 = 
\frac{\beta(k-1+\ttt)   P( k-1,\theta)}{  2\beta+\mu   }
  - \frac{\beta(k +\ttt)  P( k ,\theta)}{  2\beta+\mu }
+ \rho(\ttt) \delta_{k,\beta}
.}{difference_1}
This can be rearranged and expressed equivalently as follows:
\all{
 P( k,\theta)
 = 
\fracc{ (k-1+\ttt)   P( k-1,\theta)}{ 2+\frac{\mu}{\beta}+k+\theta }
+\frac{ 2+\frac{\mu}{\beta} }{2+\frac{\mu}{\beta}+\beta+\theta}\rho(\ttt)\delta_{k,\beta}
.}{difference_1_app}
Multiplying both sides by $2\beta+\mu$ and rearranging the terms, this can be recast as follows
\all{
 P( k,\theta)
 = 
\fracc{ (k-1+\ttt)   P( k-1,\theta)}{ 2+\frac{\mu}{\beta}+k+\theta }
+\frac{ 2+\frac{\mu}{\beta} }{2+\frac{\mu}{\beta}+\beta+\theta}\rho(\ttt)\delta_{k,\beta}
.}{difference_1_app_2}
Setting $k=\beta$, this yields $P(\beta,\theta)=\fracc{ 2+\frac{\mu}{\beta} }{2+\frac{\mu}{\beta}+\beta+\theta}~\rho(\ttt)$. For all ${k>\beta}$, the second term on the right hand side vanishes, and this equation reduces to a straightforward recursion  ${ P( k,\theta)   =  \fracc{ (k-1+\ttt)  }{ 2+\frac{\mu}{\beta}+k+\theta } P( k-1,\theta)}$,  whose solution is 
\all{&\resizebox{0.93\linewidth}{!}{$
 \DD P(k,\theta)= P(\beta,\theta) \DD \! \! \!\! \prod_{j=\beta+1}^{k}   \!\! \fracc{ (k-1+\ttt)  }{\left( 2+\fracc{\mu}{\beta}+k+\theta \right) }
= P(\beta,\theta)  \fracc{(k-1+\theta)!}{(\beta-1+\theta)!} \fracc{\Gamma \left(3+\fracc{\mu}{\beta}+\beta+\theta \right)}{\Gamma \left(3+\fracc{\mu}{\beta}+\beta+\theta \right)}$}
\nonumber \\
&=\rho(\theta) \left( 2+\frac{\mu}{\beta}\right) 
 \displaystyle \frac{\Gamma(k+\theta)}{\Gamma(\beta+\theta)}
\displaystyle \frac{\Gamma \left(\beta+2+\theta+\fracc{\mu}{\beta} \right) }
{\Gamma \left( k+3+\theta+\fracc{\mu}{\beta} \right)}  
.}{app1_p1}

\section{Obtaining the Conditional Distribution $P(\ell,\phi|k,\theta)$}\label{app:sol_2}
We begin by writing the rate equation to quantify the evolution of $N_t( k,\theta,\ell,\phi)$, which is the number of nodes with  degree $\ell$ and quality $\phi$   who are connected to a parent node of degree $k$  and quality $\theta$. Upon introduction of a new node, regardless of its quality, the following is true: if it attaches to a node of degree $\ell$ and quality $\phi$ who is the child of a parent of degree $k$ and quality $\theta$, then the degree of the receiving node increments and consequently ${N_t(k,\theta,\ell,\phi)}$ decrements. Also, ${N_t(k,\theta,\ell,\phi)}$ decrements if the new node attaches to the parent node in such  a pair of nodes. Another way that $N_t( k,\theta,\ell,\phi)$ can increment is if either there is a child-parent pair of ${(k,\theta,\ell-1,\phi)}$ or ${(k-1,\theta,\ell,\phi)}$. If the new node attaches to the child node in the former case or to the parent node in the latter case, then $N( k,\theta,\ell,\phi)$ increments. 
Finally, with probability $\rho(\phi)$, the new node will have quality $\phi$, and if the new node attaches to an existing node of degree $k-1$ and quality $\theta$, then  ${N_t(k,\theta,\ell,\phi)}$ increments. The rate equation reads
\all{
&\resizebox{0.93\linewidth}{!}{$N_{t+1}( k,\theta,\ell,\phi) - N_t( k,\theta,\ell,\phi) = \beta \left[ \fracc{(\ell-1+\phi) N_t( k,\theta,\ell-1,\phi) - (\ell+\phi) N_t( k,\theta,\ell,\phi)}{\lam+(2\beta+\mu)   t} \right]$}
\nonumber \\ &
\resizebox{0.93\linewidth}{!}{$+ \beta \left[ \fracc{(k-1+\theta) N_t( k-1,\theta,\ell,\phi) - (k+\theta) N_t( k,\theta,\ell,\phi)}{\lam+(2\beta+\mu)   t} \right]
 +\rho(\phi) \delta_{\ell,\beta} \fracc{\beta (k-1+\theta)N_t(k-1,\theta)}{\lam+(2\beta+\mu)   t}$}
}{rate_big_1}

Undertaking the same steps that let us transform~\eqref{Nkdot_1_0} into~\eqref{Pkdot_1_0}, and denoting the fraction $\frac{N( k,\theta,\ell,\phi)}{N(0)+t}$ by $n_{t}( k,\theta,\ell,\phi)$, this can be re-written in terms of $n_t(k,\theta,\ell,\phi)$ instead of $N_t(k,\theta,\ell,\phi)$. In the limit as $t \rightarrow \infty$, we can drop the $t$ subscript and obtain:
\all{
n( k,\theta,\ell,\phi) &= 
  \fracc{(\ell-1+\phi) n( k,\theta,\ell-1,\phi)}{2+\frac{\mu}{\beta}+k+\ell+\theta+\phi}
 + \fracc{(k-1+\theta) n( k-1,\theta,\ell,\phi)}{2+\frac{\mu}{\beta}+k+\ell+\theta+\phi}
 \nonumber \\ &
 +\rho(\phi) \delta_{\ell,\beta} \fracc{  (k-1+\theta)P(k-1,\theta)}{2+\frac{\mu}{\beta}+k+\ell+\theta+\phi }.
   }{difference_2}

Let us define the new sequence ${m(k,\theta,\ell,\phi)=\frac{\Gamma(3+\frac{\mu}{\beta}+k+\ell+\theta+\phi)}{(k-1+\theta)! (\ell-1+\phi)!}n(k,\theta,\ell,\phi)}$. Using this substitution and applying the properties of the Gamma function as well as the delta function, we can rewrite~\eqref{difference_2}  equivalently as
\all{
  m( k,\theta,\ell,\phi) &= 
 m( k,\theta,\ell-1,\phi)
 +m( k-1,\theta,\ell,\phi) 
 \nonumber \\ &
 +\fracc{\Gamma \left(2+\frac{\mu}{\beta}+k+\beta+\theta+\phi \right)}{(k-1+\theta)! (\beta-1+\phi)!}
 \rho(\phi) \delta_{\ell,\beta}  (k-1+\theta)P(k-1,\theta) .
   }{app2_temp1}
Using the expression in~\eqref{Pkth_fin} to rewrite the last term on the right hand side of this equation, we can express it equivalently as follows
\al{
  & m( k,\theta,\ell,\phi) = 
 m( k,\theta,\ell-1,\phi)
 +m( k-1,\theta,\ell,\phi) 
\nonumber \\ &
\resizebox{0.9\linewidth}{!}{$ + \rho(\phi)  \rho(\theta)
 \delta_{\ell,\beta}  
  \left( 2+\frac{\mu}{\beta}\right)
  \fracc{\Gamma \left(2+\frac{\mu}{\beta}+k+\beta+\theta+\phi \right)}{  (\beta-1+\theta)! (\beta-1+\phi)!}
\displaystyle \frac{\Gamma \left(\beta+2+\theta+\fracc{\mu}{\beta} \right) }
{\Gamma \left( k+2+\theta+\fracc{\mu}{\beta} \right) } $}  .
 }
Now define the generating function $\psi(z,\theta,y,\phi)=\sum_k m(k,\theta,\ell,\phi) z^{-k} y^{-\ell}$. Multiplying both sides of~\eqref{app2_temp1} by $z^{-k} y^{-\ell}$, summing over all values of $k,\ell$  and rearranging the terms, we arrive at
\all{
\psi(z,\theta,y,\phi) &= 
\fracc{\rho(\phi)  \rho(\theta)
  \left( 2+\frac{\mu}{\beta}\right) \Gamma \left(\beta+2+\theta+\fracc{\mu}{\beta} \right) }{  (\beta-1+\theta)! (\beta-1+\phi)!}
\nonumber \\ &
\times \DD \sum_{j=\beta+1}^{\infty} \fracc{\Gamma \left(2+\frac{\mu}{\beta}+j+\beta+\theta+\phi \right)}{\Gamma \left( j+2+\theta+\fracc{\mu}{\beta} \right) } 
\fracc{z^{-j} y^{-\beta}}{1-z^{-1}-y^{-1}}
.}{psi_app_1}
(The lower bound of the sum is $\beta+1$ because $P(k-1,\theta)$ is zero for $k<\beta+1$.)
The inverse transform of the factor $\frac{z^{-j} y^{-\beta}}{1-z^{-1}-y^{-1}}$in the summand  can be taken through the following steps:
\all{
\fracc{z^{-j} y^{-\beta}}{1-z^{-1}-y^{-1}} 
\xrightarrow{\mathcal{Z}^{-1}} &
\frac{1}{(2\pi i)^2}  \DD \oint \oint \fracc{z^{k-j-1} y^{\ell-\beta-1}}{1-z^{-1}-y^{-1}}   dz dy 
\nonumber \\ &
=\frac{1}{(2\pi i)^2}  \DD \oint \oint \fracc{z^{k-j} y^{\ell-\beta}}{z-\frac{y}{y-1}}   \frac{1}{y-1} dz dy 
\nonumber \\ &
= \frac{1}{(2\pi i)}  \DD \oint \oint  y^{\ell-\beta} \left(\fracc{y}{y-1} \right)^{k-j}   \frac{1}{y-1} dz dy 
\nonumber \\ &
=\frac{1}{(k-j)!} \left. \frac{d^{k-j}}{d y^{k-j}} y^{k+\ell-\beta-j}  \right|_{y=1}
= \CC{k-j+\ell-\beta}{\ell-\beta}
}{psi_inv_app}
So we can invert~\eqref{psi_app_1} term by term. We get
\all{
m(k,\theta,\ell,\phi) &= 
\fracc{\rho(\phi)  \rho(\theta)
  \left( 2+\frac{\mu}{\beta}\right) \Gamma \left(\beta+2+\theta+\fracc{\mu}{\beta} \right) }{  (\beta-1+\theta)! (\beta-1+\phi)!}
\nonumber \\ &
\times \DD \sum_{j=\beta}^{\infty} \fracc{\Gamma \left(2+\frac{\mu}{\beta}+k+\beta+\theta+\phi \right)}{\Gamma \left( k+2+\theta+\fracc{\mu}{\beta} \right) } 
 \CC{k-j+\ell-\beta}{\ell-\beta}
.}{m_fin_1_app}
From this, we readily obtain 
\all{
n(k,\theta,\ell,\phi) &= 
 \rho(\phi)  \rho(\theta) \fracc{ \Gamma \left(\beta+2+\theta+\fracc{\mu}{\beta} \right)}{\Gamma\left( 3+\frac{\mu}{\beta}+k+\ell+\theta+\phi  \right)}
\fracc{ (k-1+\theta)! (\ell-1+\phi)! }{  (\beta-1+\theta)! (\beta-1+\phi)!}
\nonumber \\ &
\times 
  \left( 2+\frac{\mu}{\beta}\right) \DD \sum_{j=\beta}^{k} \fracc{\Gamma \left(2+\frac{\mu}{\beta}+j+\beta+\theta+\phi \right)}{\Gamma \left( j+2+\theta+\fracc{\mu}{\beta} \right) } 
 \CC{k-j+\ell-\beta}{\ell-\beta}
.}{n_fin_1_app}

The last step is to abridge this quantity and the desired  NNQDD distribution, that is, $P(\ell,\phi|k,\theta)$.  Remember that the NNQDD is the fraction of $(\ell,\phi)$ nodes among the neighbors of a $(k,\theta)$ node. To obtain this fraction, we first need to obtain the total number of neighbors of $(k,\theta)$  nodes, then find the number of $(\ell,\phi)$ nodes among these nodes, and divide the latter by the former. The total number of neighbors of $(k,\theta)$ nodes is simply $k N n(k,\theta)$. The number of $(\ell,\phi)$ nodes among them equals  ${\big[n(k,\theta,\ell,\phi)+n(\ell,\phi,k,\theta)\big]N}$, because the $(\ell,\phi)$ node can both  be the parent or the child of the a $(k,\theta)$ node to be connected to it. So we have
${
P(\ell,\phi|k,\theta)= \frac{n(k,\theta,\ell,\phi)+n(\ell,\phi,k,\theta)}{k P(k,\theta)}
}$. 
Inserting the results of~\eqref{n_fin_1_app} and~\eqref{Pkth_fin} into this expression and simplifying the results, we obtain
\all{
  &P(\ell,\phi|k,\theta)=
\fracc{\rho(\phi)  }{k}
  \frac{ \Gamma \left(k+\theta+3+\fracc{\mu}{\beta} \right)}{\Gamma\left( k+\theta+3+\frac{\mu}{\beta}+\ell+\phi  \right)  }
\frac{   (\ell-1+\phi)! }{  (\beta-1+\phi)!  }  
\Gamma \left(\beta+2+\phi+\fracc{\mu}{\beta} \right)\times
\nonumber \\ 
&
\resizebox {\linewidth}{!}{$
\left[ \DD
 \sum_{j=\beta+1}^{k} \frac{\Gamma \left(j+\theta+2+\fracc{\mu}{\beta}+\beta+\phi \right) \DD \CC{k-j+\ell-\beta}{\ell-\beta}
}{\Gamma \left( j+\theta+2+\fracc{\mu}{\beta} \right)\Gamma \left(\beta+2+\phi+\fracc{\mu}{\beta} \right) } 
 +  \DD \sum_{j=\beta+1}^{\ell} \frac{\Gamma \left(j+\theta+2+\fracc{\mu}{\beta}+\beta+\phi \right)\DD \CC{\ell-j+k-\beta}{k-\beta}
}{ \Gamma \left( j+\phi+2+\fracc{\mu}{\beta} \right)\Gamma \left(\beta+2+\theta+\fracc{\mu}{\beta} \right) } 
\right].$}}{NNQDD}

\end{document}